\documentstyle[12pt,aaspp4]{article}

\begin {document}
\title{ Four nearby L dwarfs}

\author {I. Neill Reid}
\affil {Dept. of Physics \& Astronomy, University of Pennsylvania, 209 S. 33rd
Street, Philadelphia, PA 19104-6396; 
e-mail: inr@herschel.physics.upenn.edu}

\author {J. Davy Kirkpatrick}
\affil {Infrared Processing and Analysis Center, 100-22, California Institute
of Technology, Pasadena, CA 91125}

\author {J. E. Gizis}
\affil {Department of Physics and
Astronomy, University of Massachusetts, Amherst, MA 01003}
 
\author {C. C. Dahn, D. G. Monet}
\affil {U.S. Naval Observatory, P.O. Box 1149, Flagstaff, AZ 86002}

\author {Rik J. Williams}
\affil {Department of Astronomy, MSC 152, California Institute of Technology,
Pasadena, CA 91126-0152}

\author {James Liebert}
\affil {Steward Observatory, University of Arizona, Tucson, AZ 85721}

\author {A. J. Burgasser}
\affil {Dept. of Physics, 103-33,  California Institute of Technology,
Pasadena, CA 91125}

\begin{abstract}

We present spectroscopic, photometric and astrometric observations
of four bright L dwarfs
identified in the course of the 2MASS near-infrared survey. Our spectroscopic
data extend to wavelengths shortward of 5000\AA\ in the L0 dwarf 2MASSJ0746+2000
and the L4 dwarf 2MASSJ0036+1840, allowing the identification of 
absorption bands due to MgH and CaOH. The atomic resonance lines Ca I 4227\AA\
and Na I 5890/5896\AA\ are extremely strong, with the latter having an
equivalent width of 240\AA\ in the L4 dwarf. By spectral type L5, the 
D lines extend over $\sim1000$\AA\ and absorb a substantial fraction of
the flux emitted in the V band, with a corresponding effect on the (V-I)
broadband colour. The KI resonance doublet at 7665/7699\AA\ increases in equivalent
width from spectral type M3 to M7, but decreases in strength from M7 to L0 before
broadening substantially at later types. These variations are likely driven
by dust formation in these cool atmospheres.

\end{abstract}


\section {Introduction }

Familiarity with one's immediate neighbours is, in general, good policy. In the
case of the Solar Neighbourhood, our knowledge of the local constituents
forms the basis of the determination of fundamental statistical quantities
such as the luminosity function, the mass function, the local mass density and
the star formation history of the disk. Moreover, as the apparently-brightest 
members of their respective spectral classes, the nearest celestial neighbours
are most accessible to detailed astrophysical analysis. The latter
consideration is of particular importance for  objects of intrinsically low
luminosity, such as old, low-temperature white dwarfs and ultracool, very low-mass
(VLM) main-sequence dwarfs.

Until recently, the main resource for the identification of VLM dwarfs remained
the proper motion catalogues compiled by Luyten from photographic
material obtained in the 1950s and 1960s using the Palomar 48-inch Oschin
Schmidt. The development of higher-sensitivity red and photographic-infrared
emulsions in the 1970s permitted photometric surveys to extend to
somewhat larger depths, but this field of study has been revolutionised 
through the advent of deep, near-infrared all-sky surveys,
such as DENIS (Epchstein et al, 1994) and 2MASS (Skrutskie et al, 1997).
Follow-up observations of sources with extremely red JHK or optical-to-infrared
(RIJHK) colours (Delfosse et al, 1997; Kirkpatrick et al, 1999a, hereinafter
paper I) have resulted 
in the identification of numerous ultracool dwarfs. Many are spectroscopically
similar to the previously-unique white dwarf companion, GD 165B, which
has been transformed from an anomaly to a prototype. The far-red optical
spectra of these dwarfs are characterised by the disappearance of TiO and VO
absorption bands, the defining signature of spectral class M, and the
presence of metal hydride (CaH, FeH, CrH) bands and neutral alkali (Cs, 
Rb, sometimes Li) lines. The progression 
of those features was ordered in paper I to define a new spectral class, type L.

The initial sample of ultracool L dwarfs discovered by 2MASS (20 objects)
and other surveys (5 dwarfs) includes only two objects with magnitudes
brighter than K=12: Kelu 1 (Ruiz et al, 1997) and 2MASSJ1439284+192915. As
a result, apart from Ruiz et al's observations of Kelu 1, spectroscopy of these 
sources has been confined
largely to wavelengths longward of 6400\AA. We have since extended the
areal coverage of our 2MASS analysis by almost a factor of three, concentrating
on identifying late-type L dwarfs. Our current sample includes 74 
spectroscopically-confirmed L dwarfs (Kirkpatrick et al, 1999b, hereinafter Paper II). Four
(including 2MASSJ1439284+192915)
are of particular interest, since their properties imply that they lie at
distances of no more than 15 parsecs. All are sufficiently bright that
they supply an opportunity of extending high signal-to-noise observations to 
bluer wavelengths and to higher spectral resolution.
This paper provides a brief discussion of the properties of these
ultracool dwarfs.

\section {Observations}

The four L dwarfs discussed in this paper were all identified as
candidate low-temperature objects based on analysis of JHK$_S$ photometric
catalogues derived from the Two-Micron All-Sky Survey (Skrutskie et al, 
1997). 2MASSWJ1439284+192915 forms part of the original L dwarf sample
discussed in Paper I; 2MASSWJ0746425+200032 was selected
amongst a sample of candidate bright, ultracool late-type dwarfs (discussed further by
Gizis et al, 1999); 2MASSWJ0036159+182110 and 2MASSWJ1507476-162738 were
identified as likely to be mid- to late-type L dwarfs based on their
having (J-K$_S$) colours redder than 1.3 magnitudes. For brevity, we
shall refer to these sources as 2M0036, 2M0746, 2M1439 and 2M1507 throughout
the rest of this paper. The individual photometric measurements of each
object are listed in table 1: 2M0036 and 2M1507 fall in overlap
regions between separate scans and the JHK$_s$
magnitudes are averages of the two observations. A finding chart for 2M1439 is
available in Paper I, and finding charts for the other three dwarfs 
are presented in Paper II.

\subsection {Spectroscopy}

Each L dwarf has been observed using the Low Resolution Imaging Spectrograph
(Oke et al, 1995) on the Keck II telescope. Initial observations were obtained 
using a 1-arcsecond slit and the 400 l/mm grating blazed at $\lambda 8500$\AA, covering
the wavelength range 6300 to 10200\AA\ at a resolution of 9\AA. An OG570
filter was used to eliminate second-order flux. This is the standard instrumental
set-up used in our L dwarf observations, and data reduction and calibration followed the 
procedures described in paper I. The UT dates of the
individual observations were  14 \& 16 Dec 1998 (2M0036), 24 Dec 1998 (2M0746), 8 Dec 1997
(2M1439) and 24 Dec, 1998 (2M1507). 2M0746 was also observed on Dec 4, 1998 using the
modular spectrograph on the Las Campanas Observatory Du Pont 2.5-metre (see
Gizis et al, 1999 for further details), while the 2M1439 observations are described in Paper I. 
Spectral types have been derived for each dwarf based on the LRIS spectra plotted in
figure 1 following the precepts given in paper I (see Kirkpatrick et al, in prep. for
further details).

We have supplemented these intermediate-resolution red spectra  with a range of other 
observations. 

{\sl LRIS: blue spectra}

We also have shorter-wavelength LRIS observations of 2M0036, 2M0746 and 2M1507, using the 300 l/mm
grating blazed at 5000\AA. Those
spectra were obtained on 25 Dec, 1998 (2M0036), 5 March, 1999 (2M0746) and 17 July, 1999.
The respective exposure times were 1800, 1800 and 3600 seconds respectively. 
As with the standard far-red observations, we used a 
1-arcsecond slit, providing a spectral resolution of $\sim6\AA$ and wavelength
coverage from $\sim3900$ to 7800\AA. No order-sorting filters were
employed. The data reduction procedures mirror those used in analysing the
red data, with flux calibration provided through observations of
Hiltner 600 and LTT 9491(Hamuy et al, 1994).  
We also obtained lower-resolution data covering the 5400 to 10400\AA\ region for 2M1507
using a 158 l/mm grating on the red channel of double spectrograph on the Hale 200-inc (5.08 metre)
telescope. Those data are consistent with the higher signal-to-noise Keck spectrum.

Figure 2 plots the reduced spectra, where we include, for comparison, our observations of the
late-type M dwarf BRI0021-0214 (M9.5, Kirkpatrick et al, 1995) and data for
the L2 dwarf, Kelu 1 (Ruiz et al, 1997). The latter spectrum was kindly made
available by S. Leggett. Figure 3 provides an
expanded view of the 4500 to 6600\AA\ region. The more prominent molecular and atomic
features are labelled in both figures. Note, in particular, the increasing strength of
both the potassium 7665/7699 and sodium 5890/5896 resonance doublets as one progresses
from spectral type M9.5 to L5.

{\sl HIRES observations}

Finally, we have obtained higher-resolution echelle spectra of
all four L dwarfs using HIRES (Vogt et al, 1994) on the Keck I telescope. 
The observations were obtained 24 Aug 1998 (2M0036, 2M1439), 6 March 1999
(2M0746) and 14 June, 1999 (2M1507). In each case, the data provide
partial coverage of the wavelength range $\lambda\lambda 6000 - 8500$\AA,
including important features such as Li I 6708\AA, K I 7665/7699\AA, Rb I
7800 \& 7948\AA, and the Na I 8183/8195\AA\ doublet. Total exposure times of
6000 seconds were accrued on each source. As discussed
further below, lithium was not detected in any of these four L dwarfs. 

The HIRES data were flat-field corrected and the spectra extracted
using programmes written by T. Barlow. The wavelength calibration,
based on Th-Ar arc lamp exposures, was determined using the iraf
routines ECIDENTIFY and DISPCOR. We have not attempted to set
these data on a flux scale. Radial velocities were computed for
each star either from the measured wavelength of the H$\alpha$ emission
line (in 2M0746 and 2M1439) or by measuring the central wavelengths of atomic
lines due to Cs and Rb, adopting heliocentric corrections
given by the IRAF RV package. Our radial velocity measurements for
M dwarfs from the Marcy \& Benitz (1989) sample indicate that the
latter technique can give velocities accurate to $\pm$1.5 kms$^{-1}$.
However, the atomic lines are relatively broad in the L dwarfs, and an
internal comparison of the individual measurements suggests that the
uncertainty is 2-3 kms$^{-1}$. Save for 2M1439, the uncertainties
in the derived space motions are dominated by the parallax measurements.

\subsection { Photometry}

CCD images in several passbands have been obtained of all four 
L dwarfs discussed in this paper. The observations were made 
using the 40-inch telescope at the Flagstaff station of
the US Naval Observatory. Full details of the data reduction
and calibration process are given by Dahn et al (in prep.). Those
data are listed in Table 1.

In addition to these direct measurements, we have used the calibrated
spectra plotted in figures 1 and 2 to synthesise (B-V), (V-R) 
and (V-I) colours. As in Paper I, square passbands
are adopted for each filter, and the flux zeropoints are those of
the Johnson/Kron-Cousins system (Bessell, 1979). In general, there
is reasonable  agreement between the spectroscopic colours and
the available direct measurements. 

Finally, we have estimated bolometric
magnitudes for each dwarf. While none of these sources, and relatively
few late-type M or L dwarfs in general, have observations at wavelengths
longward of 2.2$\mu m$, the available data suggest that m$_{bol}$ can be
inferred with reasonable accuracy from the observed magnitude at the 1.25$\mu m$
J band. Leggett et al (1996) infer BC$_J = 2.07$ magnitudes for the 
M6.5 dwarf GJ 1111; Tinney et al (1993) infer BC$_J =  1.7$ mag for the L4
dwarf GD165B; and Leggett et al (1999) have derived BC$_J = 2.19$ mag for
Gl 229B. These results indicate that there is relatively little variation
in BC$_J$ over this temperature range ($\sim2700$K to $\sim950$K) and we have
adopted a uniform correction of M$_{bol}$=M$_J$ + 1.75 mag for each object in
the current sample. 

\subsection {Astrometry}

All four L dwarfs discussed in this paper have been placed on the US Naval 
Observatory (Flagstaff) CCD parallax programme (Monet et al, 1992). Preliminary
absolute parallaxes are available in each case, and those data
are listed in Table 2. These observations are also used to 
derive absolute proper motions, and the results for 2M0036, 2M0746 and 2M1439 are
listed in Table 2. In the case of 2M1507, the USNO observations
span a period of only 107 days, leading to significant uncertainties in $\mu$ 
and $\theta$. Fortunately, that dwarf is visible on both the first and
second epoch plates taken by the UK Schmidt telescope as part of the 
southern sky survey. Most of the L dwarfs listed in paper I
are visible on the POSS II IVN I-band plates, and several are also
detected on POSS II IIIaF (R-band) plate material. 2M1507 is
unusual in that it is sufficiently bright to be detected on even the 
IIIaJ (blue-green) 1st-epoch UKST plates. The time difference between the two
UKST observations is 11 years, sufficient to allow a more accurate
estimate of the proper motion than provided by current CCD observations.
We have used standard profile-fitting techniques
to measure the displacement between the two epochs and derive an annual proper motion
close to 1 arcsecond directed almost due south. Note that three of the four L
dwarfs have motions consistent with their inclusion in the Luyten Half Second catalogue.

\section {Discussion}

These new observations allow us to investigate further the spectral energy distribution
and atmospheric composition of L dwarfs. In addition, we can determine
space velocities for the four objects in the present sample.

\subsection {Molecular features}

Far-red optical spectra of L dwarfs show that metal hydride bands, notably CaH, 
FeH and CrH, become increasingly prominent with decreasing temperature (later
spectral types). This behaviour is reminiscent of that observed in late-type
metal-poor subdwarfs. In both cases, the greater visibility of the hydride
bands reflects decreasing strength of TiO and VO absorption, albeit governed
by two different mechanisms: in the subdwarfs, the weak oxides are due
to an overall scarcity of metals; in the L dwarfs, TiO and VO are depleted
as dust particles, mainly perovskite, CaTiO$_3$, and solid-phase VO respectively.
Other minerals, such as enstatite (MgSi$_3$) and forsterite (Mg$_2$SiO$_4$), are
also expected to condense at temperatures between $\sim2100$K and 
1500K (Fegley \& Lodders, 1996;
Burrows \& Sharp, 1999; Lodders, 1999).

By analogy with cool subdwarfs, other hydrides are expected to be visible
at shorter wavelengths - in particular, MgH (Cottrell, 1978). Figure 4 plots
our LRIS observations of three extreme ([m/H]$< -1.5$) dwarfs: LHS 489 (esdM0
on the system defined by Gizis, 1997), 
LHS 453 (esdM3.5) and LHS 375 (esdM5). Those observations were obtained on 17 July, 1999
using the same instrumental setup and data reduction process as in our observations
of 2M1507. Comparing spectra for the two sets of objects reveals both significant
similarities and differences. The L dwarfs are substantially cooler than the 
$\sim 3000$ to 4000K esdMs, leading to  much steeper spectral energy distributions
in the former than the latter. In both cases, however, the most prominent molecular
absorption is due to metal hydrides, 
with the 5200\AA\ MgH feature obvious in all of the L dwarfs.
The $\lambda4788$\AA\ band is clearly present
in the L5, 2M1507, and is barely detected in 2M0036.

TiO bands at 4761, 4954 and 5448\AA\ are
evident in 2M0746, but have disappeared by spectral type L4. All of the L dwarfs
also  exhibit strong absorption at $\sim5500$\AA\, with the band most
prominent in 2M0746. This feature is likely to be 
calcium hydroxide, CaOH, originally identified
in mid-type M dwarfs by Pesch (1972) and increasingly strong in
later-type M dwarfs. This molecule also contributes a diffuse
band at $\sim6230$\AA\ (Pearse \& Gaydon, 1965), which is blended with the 
$\lambda6158$\AA\ $\gamma'$ TiO
band in M dwarfs (Boeshaar, 1976). The latter two bands are likely responsible for the 
substantial, double-bottomed absorption feature at $\sim6200$\AA\ in 2M0746, evident
as a shallower depression in the L4, 2M0036. Both CaOH and MgH can be identified in
the spectrum of Kelu 1 presented by Ruiz et al (1997).
Finally, the
VO $\lambda 5736$\AA\ band is probably responsible for  a
relatively weak absorption feature in 2M0746.

\subsection {Atomic lines}

Table 3 lists equivalent widths for some of the more prominent
atomic lines present in the spectra of these objects. We list 
results from measurements of both the LRIS spectra plotted in figure
1 and of our HIRES data. The latter provide only
incomplete wavelength coverage, but the higher-resolution data allow
more accurate measurements of weaker lines. In particular, the Ca I
6572\AA\ and 8256\AA\ absorption lines and H$\alpha$ emission
are barely detectable in the LRIS spectra, where the measured equivalent
widths have a 1$\sigma$ uncertainty of $\pm0.5$\AA. 

One of the strongest features, either atomic or molecular,
in the far-red spectra of L dwarfs is the
K I 7665/7699\AA\ resonance doublet.  Those lines have individual equivalent
widths of 10 to 12 \AA\ at spectral type L0, but increase dramatically in
strength with decreasing temperature to the extent that the lines effectively merge
at $\sim$L5, where the composite feature has a width exceeding 100\AA. Similarly,
the Rb I and Cs I lines show distinctly non-linear behaviour, increasing
substantially in strength between the L4 dwarf 2M0036 and 2M1507 (L5).

In paper I we proposed that this behaviour is another
consequence of dust formation. 
As discussed further in section 3.5, 
dust initially contributes a scattering layer at late-type M dwarfs,
but in lower temperature atmospheres (later spectral types) the dust
particles either `rain out' to greater depths (below the photosphere)
or form larger particles,
in either case reducing scattering at optical wavelengths. The overall atmospheric
transparency is further increased as metals are transformed to solid
phase, both by the removal of TiO and VO molecular absorption, and through
the scarcity of free electrons and the resulting reduced
level of H$^-$ continuum opacity.

The $\tau=1$ photosphere lies at a large physical
depth within the low opacity L-dwarf atmosphere, with the result that
the column density of (relatively) undepleted elements, such as the alkali metals,  can
reach very substantial values. In addition, gas pressure increases with
increasing depth leading to substantial van der Waal's broadening, as in
degenerate white dwarfs. Both effects lead to strong atomic lines.
As discussed in paper I, the relative strengths of the resonance lines of those
species visible in the far red (K, Cs, Rb) are consistent with their
relative abundances in the Sun. (The Ca I 6572\AA\ line and the Na I 8183/8194
doublet are higher-order transitions.)
Sodium is not expected to
form grains until temperatures of less than 1200K (the T-dwarf r\'egime)
and, with a higher abundance than potassium ([Na] = 6.31 as compared to
[K]=5.13 for [H]=12.0, where [m] is the logarithmic abundance), the 
D lines at 5890, 5896\AA\ are predicted  to grow in strength at
earlier spectral types than the K I doublet.

This prediction is confirmed by the spectra plotted in figures 2
and 3. The sodium lines, which already have the
substantial equivalent width of $\sim36$\AA\ in the M9.5 BRI0021 have
doubled in strength to $\sim80$\AA\ by spectral type L0.5 (2M0746).
We measure an equivalent width of $\sim170$\AA\ in the
L2 dwarf Kelu 1, and our spectrum of 2M0036 yields an equivalent width 
of $\sim240$\AA\ for that L3.5 dwarf, although identifying appropriate 
pseudo-continuum points is becoming problematic at these later spectral types.

Initial observations of 2M1507 with the Palomar double spectrograph
revealed a steeply declining spectrum shortward of 6700\AA, with
no significant flux detected shortward of $\sim6000$\AA. Our surmise
that this might reflect increased sodium absorption is confirmed
spectacularly by the LRIS data plotted in figure 2. Superimposed on
the steeply-rising underlying spectrum, the D lines produce a smooth,
concave feature spanning over 1500\AA, with MgH the only identifiable
absorption feature between 4500 and 6500\AA. 
The blue wing of this atomic doublet extends to $\sim5000$\AA,
where the spectral energy distribution reaches a mild peak before
declining towards shorter wavelengths. The red wing of the Ca I
4227\AA\ resonance line probably contributes to that smooth decline.
Similar behaviour in the K I 7665/7699 doublet at 
much cooler temperatures is partly responsible for the steep
flux gradient between 8000 and 9000\AA\ in the energy distribution of methane-rich T dwarfs 
such as Gl 229B (Oppenheimer et al, 1998). 

\subsection {Chromospheric activity and lithium absorption}

H$\alpha$ emission has long been known as an indicator of chromospheric
activity amongst M dwarfs, and earlier studies suggested that emission became
increasingly common amongst later spectral types. 
Gizis et al (1999), however, have re-examined the distribution
of chromospheric activity as a function of spectral types, using 
2MASS observations to define a photometrically-selected sample of M dwarfs
which includes a significantly larger number of ultracool ($>$M7)
objects than was previously available. Analysis of that sampel shows that the
frequency of H$\alpha$ emission peaks at close to 100\% at spectral
type M7 and declines thereafter. Only 45\%
of known early-type ($\le$L3) L dwarfs  have emission lines with equivalent
widths exceeding $\sim2$\AA\, while none of the later-type L dwarfs in paper I
have detectable emission, despite the low continuum flux in the latter objects.

The four sources considered here show behaviour similar to the dwarfs
in the paper I sample.
Both of the earlier-type dwarfs have weak H$\alpha$ emission, while no emission
is detectable in the two later-type dwarfs. Figure 5 plots our HIRES data for
the H$\alpha$ region of the spectrum in three objects - the 2M1507 data are
of low signal to noise at these wavelengths and essentially featureless. 
Both of the H$\alpha$ profiles, but oparticularly 2M1439, 
appear to have a narrow core centred on a broader pedestal. This is
morphology is also found in approximately 10\% of the ultracool M dwarfs.

None of these dwarfs
is particularly active. We can use our flux-calibrated  LRIS
spectra to determine emission line fluxes from our measured equivalent widths.
In the case of 2M0746, we derive F$_\lambda \sim 2.6 \times 10^{-16}$
 erg cm$^{-2}$ sec$^{-1}$, while
for 2M1439 we find F$_\lambda \sim 7.8 \times 10^{-17}$ erg cm$^{-2}$ sec$^{-1}$.
These correspond to activity ratios, L$_\alpha$/L$_{bol}$, of 10$^{-5.5}$ and 10$^{-5.4}$
respectively, values which are almost two orders of magnitude lower than the
typical level of activity amongst M dwarfs,
$\langle {L_\alpha \over L_{bol}} \rangle \sim 10^{-3.8}$
(Hawley et al, 1996; Gizis et al, 1999) and an order of magnitude
below the quiescent state of the ultracool M9.5e dwarf, 2MASSW J0149090+295613
$\langle {L_\alpha \over L_{bol}} \rangle \sim 10^{-4.6}$
(Liebert et al, 1999).
The upper limits corresponding to
non-detection imply even lower activity ratios for the two later-type dwarfs.

Our HIRES observations also allow us to set limits on the equivalent width
of the Li I 6708\AA\ absorption line
in these four dwarfs. The presence of
atmospheric lithium in late-type dwarfs is now well-recognised  as
an indicator of substellar mass (Rebolo et al, 1992; Magazzu et al, 1993). Recent models 
indicate that all dwarfs with masses exceeding 0.06M$_\odot$ should
have depleted lithium by the time that their surface temperature has fallen to $\sim2400$K, 
equivalent to spectral type M7 (Baraffe et al, 1998). Based on the
scale derived in paper I, we estimate temperatures between $\sim2100$K (2M0746)
and $\sim$1700K (2M1507) for the four dwarfs considered here. None has 
lithium absorption exceeding 200m\AA.  Approximately one in
four of the 80 L dwarfs identified to date from 2MASS data have 
detectable lithium absorption, with equivalent widths rising to $\sim20$\AA\
amongst the later spectral types (Kirkpatrick et al, in prep). Thus, the
absence of detectable lithium in these dwarfs implies that all
of these dwarfs have depleted their primordial store of lithium: that is, all
four have 
masses which exceed 0.06M$_\odot$.

That these four L dwarfs have masses relatviely close to the hydrogen-burning
limit is not surprising. Figure 19 in paper I shows the predicted time
evolution of temperature for models spanning the mass range 0.01
to 0.1 M$_\odot$. Both the calculations by Baraffe et al (1998) and Burrows et al (1997)
predict that objects with masses as high as $\sim0.08 M_\odot$ (i.e. very
low mass stars) can acheive temperatures of 2100K, the value we 
associate with spectral class L0. Similarly, the
upper mass limit at T$_{eff} \sim 1700$K (L5) is 0.07 to 0.075 M$_\odot$.
In both cases, higher mass objects spend several Gyrs at those temperatures, while
brown dwarfs with masses below $\sim0.06 M_\odot$ have total residence times
of no more than $\sim10^8$ years. Those circumstances lead to 
much higher probabilities of detecting high-mass brown dwarfs
and very low mass stars at early and mid-L spectral types. Lower-mass
brown dwarfs make a larger  contribution to samples of late L dwarfs.

\subsection {Kinematics}

Table 2 shows that all four L dwarfs have substantial velocities relative
to the Sun. 
The correlation between space motion is statistical rather than direct, but
since velocity dispersion increases with age, there is less ambiguity in
interpreting a high velocity as implying a relatively old age than in
taking a low velocity as implying youth. Representative tracers
of the `young disk' population (A stars, active late-type dwarfs, Cepheids)
indicate that a 1-Gyr old population can be modelled as a Schwarzschild ellipsoid with
[U = -10, V=-10, W=-7; $\sigma_U = 38 $, $\sigma_V = 26 $,
$\sigma_W = 21$ kms$^{-1}$] (Soderblom, 1990). The average space velocity
for those kinematics is V$_{tot} = 34$kms$^{-1}$, and even the lowest velocity
L dwarf in the current sample, 2M1507, lies at the 75th percentile of the predicted
velocity distribution, albeit within 1$\sigma$ of the mean.

The measured velocities are more characteristic of an older stellar population.
Hawley et al (1996) derive the following ellipsoid from observations of nearby M
dwarfs: 
[U = -10, V = -21 , W = -8 ; $\sigma_U = 38 $, $\sigma_V = 26 $,
$\sigma_W = 21$kms$^{-1}$]. Matched against that distribution, 2M0036, 2M0746, 2M1439
and 2M1507 fall at the 48th, 68th, 93rd and 39th percentiles. It therefore 
seems unlikely that these dwarfs are younger than $\sim1$ Gyr, further corroborating
their identification as high-mass brown dwarfs
or low-mass stars.

\subsection {Colour-magnitude diagrams}

Our spectrophotometry provides the first opportunity of examining
the BV colours of L dwarfs. It is notable that the (B-V) colour
inferred from the spectrophotometry for 2M0036 is $\sim0.5$ magnitudes
bluer than that for the L0 dwarf, 2M0746. This counter-intuitive 
blueward evolution with decreasing temperature can be ascribed
in large part to the increasing strength of the Na D lines. A complementary
effect can be expected in the (V-I) colour.

Figure 6 plots the (M$_V$, (V-I)) colour magnitude diagram for nearby
stars with accurate parallaxes and reliable photometry 
(Bessell, 1990; Leggett, 1992) supplemented by our own
own data for the four bright L dwarfs
discussed in this paper. 2M1507 has a formal visual absolute magnitude
of M$_V \sim 22.9$. Among M dwarfs, (V-I) reaches a local maximum at
spectral type $\approx$M7: VB8 (M7) has(V-I)=4.56 mag, while VB10 is
only slightly redder at (V-I)$\sim 4.7$ mag. 
Later-type M dwarfs, such as
LHS 2924 (M9, (V-I)$\sim 4.37$), have lower luminosities, but bluer (V-I)
colours (Monet et al, 1992). Our new data show that the march redward resumes amongst the L dwarfs,
with the growing strength of the sodium D lines contributing to
the decreased flux in the V band, notably the nearly 1 magnitude offset in
(V-I) between 2M0036 (L4) and 2M1507 (L5). 

The cause of the reversal in (V-I) colour amongst the later-type M
dwarfs has received little discussion in the literature.  Spectroscopy
shows no evidence for increased molecular absorption in the far red, which
might decrease the emergent flux in the I band. Indeed, the strongest
molecular absorber, TiO,
peaks between $\sim$M6 and M8 (the $\gamma$ 7050\AA\ band is strongest at
M6.5) and decreases in strength in dwarfs of later spectral type, while
other species, such as VO, have less extensive absorption bands. 
These same stars show a near-monotonic trend toward redder colours with
decreasing luminosity in optical to infrared colours, such as (I-J) (figure 4).
This suggests that the colour reversal in (V-I) stems primarily from increased flux
in the V band rather than a deficit at I-band\footnote{Note, however, that the growth
in strength of the K I 7665/7699\AA\ resonance lines amongst the later L dwarfs
is likely to result in an effect on M$_I$ analagous to the effect
of the D lines on M$_V$ between spectral types L4 and L5. 
Gl 229B is almost 1 magnitude redder in (I-J) than the L8 dwarf 2MASSWJ1632291+190441.}.
The (M$_J$, (I-J)) diagram beautifully illustrates the 'step'
in the main sequence at spectral type $\approx$M4, originally
highlighted by Reid \& Gizis (1997) and probably due to
the onset of convection. The L0 dwarf 2M0746 lies $\sim0.7$ magnitudes above
the `main sequence' in this plane, raising the possibility that it is an
equal-mass binary.

We suggest that the behaviour the (V-I) colour is driven by the formation of dust
in the upper atmospheric layers of mid-type M dwarfs, and by the subsequent
evolution of the particle size and/or spatial distribution at lower effective
temperatures. Tsuji et al (1996) originally demonstrated that dust
formation has an important effect on the emergent spectral energy distribution of
cool dwarfs, notably a reduction in the strength of the near-infrared H$_2$O bands
due to atmospheric heating through dust re-radiation. Allowing for the 
latter effect reconciles a long-standing discrepancy between theoretical
models and observations of late-type M dwarfs (cf Reid \& Gilmore, 1984). Our
hypothesis is that the colour reversal in (V-I) has the same origin.

Tsuji et al (1996) place the onset of dust formation at T$_{eff} \sim 2600$K.
Leggett et al (1996) estimate T$_{eff} \sim 2700$K for the M6.5 dwarf GJ 1111,
suggesting that dust should become evident at spectral types of $\approx$M7 and 
later. Supporting evidence for
dust formation at this spectral type comes from variations in the
equivalent width of the 7665/7699 KI doublet in mid- the late-M dwarfs. Figure 7
plots HIRES data covering this region of the spectrum for eight dwarfs
with spectral types between M3 and L4. While the detailed profile of the
shorter wavelength component is obscured partially by terrestial O$_2$
absorption (the A band), it is clear that the overall variation mimics that
of the (V-I) colour. The equivalent widths rise to a maximum at spectral type
M6.5/M7, declines noticeably in strength to spectral type M9.5/L0, before
increasing dramatically throughout the L dwarf sequence, as discussed above and
in paper I.

We explain this behaviour as a combination of two effects. First, at spectral
types M7-M9.5, dust is present in the atmosphere in sufficient quantities to
act as a scattering layer, raising the atmospheric opacity and hence reducing the
physical depth (and hence both gas pressure and column density) of the $\tau=1$
layer for line formation; second, dust re-radiation not only reduces the strength
of the H$_2$O bands, but also increases the flux emitted at visual 
wavelengths, resulting in bluer (V-I) colours. Dust formation may 
also reduce the overall molecular (mainly TiO) opacity to a greater
extent at visual wavelengths than at 0.8$\mu m$. In late M dwarfs, such as LHS 2924, 
the total flux emitted at visual wavelengths amounts to less than 0.1\% of the
bolometric flux, so a small flux redistribution can have a large effect on F$_V$.
Section 3.2 summarises the likely explanations for
the increased equivalent widths in all of the alkali lines 
at spectral types beyond L0: increased particle size or rain out. More detailed
spectrophotometry of mid- to late-type M dwarfs at blue and visual wavelengths 
can test the overall validity of this hypothesis. 

\section {Summary and Conclusions}

We have presented spectroscopic and photometric data for four
bright L dwarfs lying at distances of less than 15 parsecs from the
Sun. Our observations permit the first detailed examination of the 
properties of these objects at blue and visual wavelengths, revealing the
presence of MgH and CaOH molecular absorption. In addition, the sodium 
D lines are extremely strong, reaching equivalent widths in excess of
240\AA\ in later-type L dwarfs. This behaviour likely stems from the
low atmospheric opacity in the latter objects and the consequent
substantial pressure broadening. The growth in strength of the Na D 
lines is also responsible for the (V-I) colour
becoming significantly redder between spectral types L4 and L5. The
KI 7665/7699 doublet probably has a similar effect on the I-band
flux between spectral types L8+ and T.

Dust formation is clearly an important factor governing spectral
evolution at these low temperatures. Theoretical models suggest
that dust first forms, primarily as TiO-based agglomerates, at $\sim2600$K, a 
prediction which is supported by the behaviour of the KI lines at
7665/7699\AA\ at spectral types between M3
and L0. Indeed, we suggest that the reversal in the (M$_V$, (V-I))
relation at spectral type $\approx$M7 may be a consequence of both
lower molecular opacities and 
dust re-radiation heating the atmosphere, with a consequent increase
in the flux emitted at visual wavelengths.

None of the four L dwarfs considered here has detectable lithium
absorption, indicating masses of at least 0.06M$_\odot$. All, however, are
also chromospherically inactive, implying masses close to, if not
below, the hydrogen-burning limit, and the relatively high space motions
suggest ages of $\sim1$ Gyr or more. Taken together, these indicators
suggest masses of from 0.07 to 0.09M$_\odot$. Further detailed observations
of these and other bright L dwarfs will prove important in determining
the general physical characteristics of these objects. 

\acknowledgements
{The authors would like to thank Pat Boeshaar for illuminating discussion
on the CaOH molecule and Sandy Legget for providing a copy of the blue
spectrum of Kelu 1. We would also like to thank the staff of the
Keck Observatories for their skilled and enthusiastic support in acquiring the
observations for this project. \\
JDK, INR and JL acknowledge funding through a NASA/JPL grant to 2MASS Core Project
science.  AJB and RJW acknowledge support from this grant.\\
Much of the V,I photometry
  reported in Sec. 2.2 was obtained by H. Harris as part of the USNO
  parallax efforts and we thank him for allowing the use of it here.
  The astrometric data reported in Sec. 2.3 were aquired by a team of
  observers which includes B. Canzian, H. Guetter, S. Levine, C. Luginbuhl,
  A. Monet, R. Stone, and R. Walker.  We thank them for their contributions. \\
This publication makes use of data from the 2-Micron All-Sky Survey, which is
a joint project of the University of Massachusetts and the Infrared Processing and
Analysis Center, funded by the National Aeronautics and Space Administration and the
National Science Foundation. \\
 The Keck Observatory
is operated by the Californian Association for Research in Astronomy, and
was made possible by generous grants from the Keck W. M. Foundation. \\ 
 This work is based partly on photographic plates obtained at the Palomar
Observatory 48-inch Oschin Telescope for the Second Palomar
Observatory Sky Survey which was funded by the Eastman Kodak
Company, the National Geographic Society, the Samuel Oschin
Foundation, the Alfred Sloan Foundation, the National Science
Foundation grants AST84-08225, AST87-19465, AST90-23115 and
AST93-18984,  and the National Aeronautics and Space Administration 
grants NGL 05002140 and NAGW 1710. JDK and AJB acknowledge the support of
the Jet Propulsion Laboratory, California Institute of Technology, which is
operated under contract with the National Aeronautics and Space Administration.}

{}
\clearpage
\begin{table}
\begin{center}
{\bf Table 1} \\
{ L dwarf photometry}
\begin{tabular}{lcrrrrrrrr}
\tableline\tableline
Name & Sp. & (B-V)$_{SP}$ & (V-R)$_{SP}$ & V & I$_C$ & J & H & K \\
\tableline
2M0036 & L3.5 & 1.7$\pm$0.2 & 3.0$\pm$0.1 & 21.33$\pm.06$ & 16.10$\pm.02$ & 12.44 & 11.58 & 11.03\\
2M0746 & L0.5 & 2.1$\pm0.2$ & 2.3$\pm0.1$ & 19.87$\pm.06$ & 15.11$\pm.02$ & 11.74 & 11.00 & 10.49 \\
2M1439 & L1 & & & 21.04$\pm.02$& 16.12$\pm.02$ & 12.76 & 12.05 & 11.58 \\
2M1507 & L5 & & & 22.9$\pm.5$ & 16.65$\pm.02$ & 12.82 & 11.90 & 11.30 \\
\tableline\tableline
\end{tabular}
\end{center}
\end{table}
\clearpage
\begin{table}
\begin{center}
{\bf Table 2} \\
{ L dwarf astrometry and kinematics}
\begin{tabular}{lcccc}
\tableline\tableline
Parameter & 2M0036 & 2M0746 & 2M1439 & 2M1507 \\
\tableline
$\mu$ arcsec yr$^{-1}$& 0.833$\pm0.073$ & 0.464$\pm0.091$ & 1.2943$\pm0.0012$ & 0.99$\pm.05$ \\
$\theta$ degrees & 80$\pm15$ & 250$\pm2$ & 288.3$\pm0.1$ & 174$\pm5$ \\
$\pi$ milliarcsec & 92.2$\pm$16.3& 69.4$\pm16$ & 69.5$\pm0.6$ & 117.5$\pm 25.2$ \\
V$_{rad}$ kms$^{-1}$ & 21.7$\pm3.0$ & 56.6$\pm2.0$ & -23.9$\pm2.0$ &-36.4$\pm3.0$ \\
M$_K$ & 10.86$\pm0.35$ & 9.70$\pm0.45$ & 10.70 $\pm0.02$ & 11.65$\pm0.42$ \\
M$_{bol}$ & 14.00 & 12.70 & 13.65 & 14.90 \\
U kms$^{-1}$ & -46.3 & -60.6 & -80.7 & -15.8 \\
V kms$^{-1}$ & -4.1 & -21.5 & -39.1 & -17.5 \\
W kms$^{-1}$ & -11.9& -8.8 & 17.9 & -48.6\\
V$_{tot}$ kms$^{-1}$ & 48.0$\pm8.5$ & 64.9$\pm13.9$ & 91.4$\pm2.5$ & 54.0$\pm9.5$\\
\tableline\tableline
\end{tabular}
\end{center}
\end{table}
\clearpage
\begin{table}
\begin{center}
{\bf Table 3} \\
{ Equivalent widths for atomic lines}

\begin{tabular}{lcccc}
\tableline\tableline
 & 2M0746 & 2M1439 & 2M0036 & 2M1507 \\
 &  L0.5 & L1 & L3.5 & L5 \\
\tableline
 & LRIS &\\
K I 7665 & 9.6$\pm0.5$\AA\ & 19$\pm0.5$\AA\ & $<$85\AA\ $>$150\AA\ \\
K I 7699 & 8.8 & 9.7 & \\
Rb I 7800 & 2.0 & 3.0 & 3.3 & 6.7 \\
Rb I 7947 & 3.1 & 3.0 & 4.0 & 7.6 \\
Na I 8183/94 & 9.1 & 9.9 & 7.5 & 5.1 \\
Cs I 8521 & 2.7 & 2.1 &2.5 & 4.6 \\
Cs I 8943 & 2.2 & 1.7: & 3.6 & 3.5 \\
\tableline
 & HIRES &\\
H$\alpha$ &1.38$\pm0.05$ & 1.13$\pm0.05$& \\
Ca I 6572 & 0.69 & 0.55 & 0.86 \\
Li I 6708 & $<$0.2 & $<0.05$ & $<$0.1 & $<0.1$ \\
Rb I 7800 & & &4.32  \\
Rb I 7947 & & 2.0 & 3.59 &4.3 \\
Na I 8183 & 1.58 &  &2.04 & 1.39 \\
Na I 8194 & 3.26 &  & 3.76& 1.82 \\
8256 & & 0.57 & 0.57& \\
Cs I 8521& & 1.14 & 2.15& 3.44 \\
\tableline\tableline
\end{tabular}
\end{center}
\end{table}
\clearpage

\centerline{FIGURE CAPTIONS}
\vskip2em

\figcaption{Far-red optical spectra of the four bright L dwarfs discussed in this paper.}

\figcaption{Blue/visual LRIS spectra of the 4000 to 7800\AA\ region in late-type dwarfs.
In addition to three of the L dwrafs discussed is this paper, we plot spectra for
the M9.5 dwarf BRI0021-0214 and for 2MASSWJ150654.4+131206, a bright early-type L dwarf.
The more prominent features are identified.}

\figcaption{Expanded reproductions of the 4500 to 6600\AA\ region of the
spectrum for the late-type dwarfs plotted in figure 2, highlighting
the strong MgH bands and the substantial increase in strength of the Na D lines.}

\figcaption {Blue-green spectra of three extreme subdwarfs.}

\figcaption {The H$\alpha$ region in 2M0746, 2M1439 and 2M0036  from our
HIRES observations. The Ca I 6572 absorption is also evident in these
spectra, as are TiO bands in the earlier-type dwarfs.}

\figcaption {The (M$_V$), (V-I)) and (M$_J$, (I-J)) diagrams defined by
nearby stars with  accurate
parallax measurements. Crosses mark objects with Hipparcos
astrometry; open circles are stars in the 8-parsec sample (Reid \& Gizis, 1997) or
with ground-based parallax measurements by Monet et al (1992) or
Tinney (1996). Note the reversal in (V-I) colour at M$_V > 18$.
The four L dwarfs discussed in the present paper are
plotted as solid points.}

\figcaption {High-resolution spectra of the K I 7665/7699 doublet in
dwarfs with spectral types between M3 and L4. VB 8 and, to a lesser extent,
VB 10 both exhibit chromospheric reversals in the core of both lines. }

\setcounter{figure} {0}
\begin{figure}
\plotone{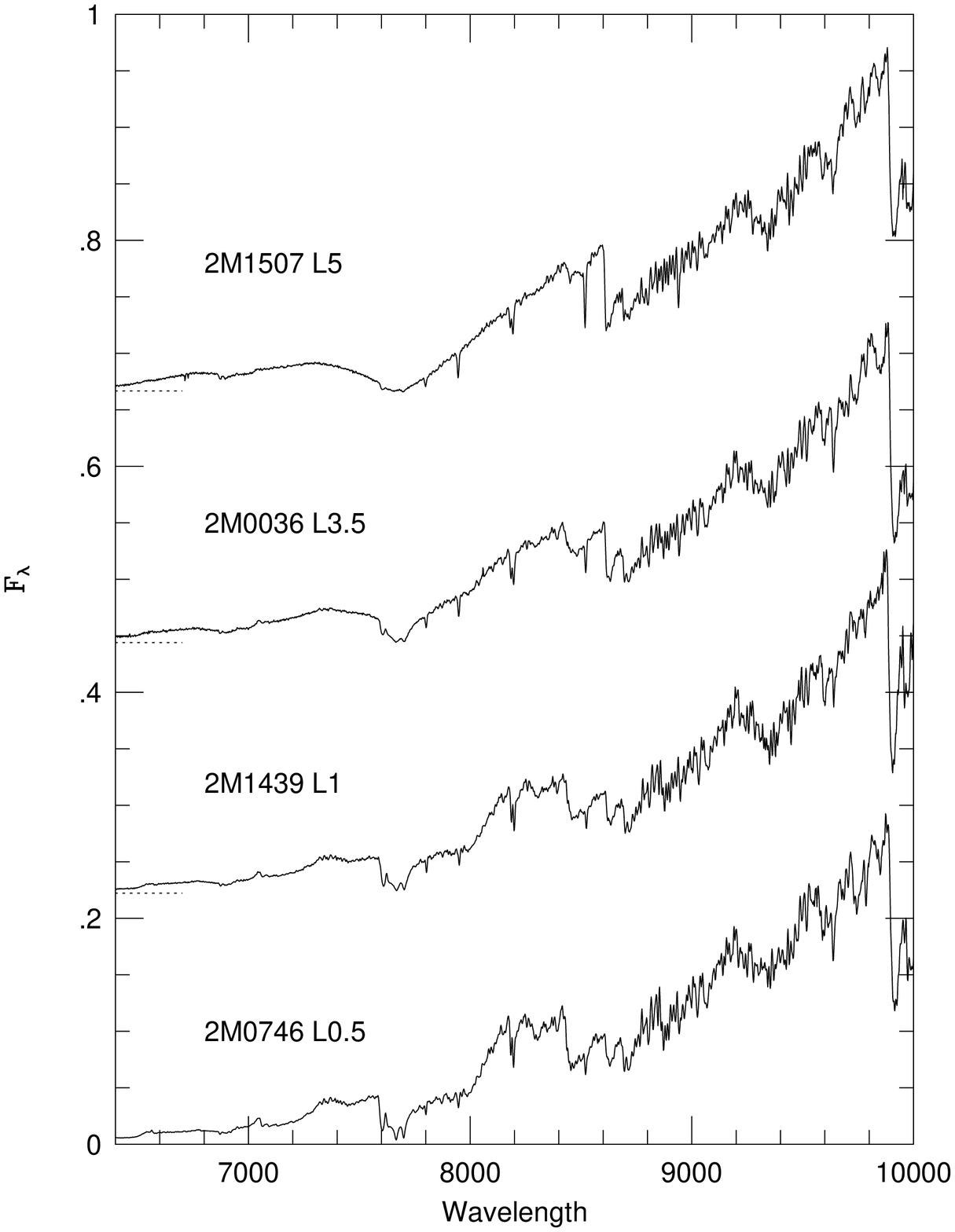}
\caption{}
\end{figure}

\begin{figure}
\plotone{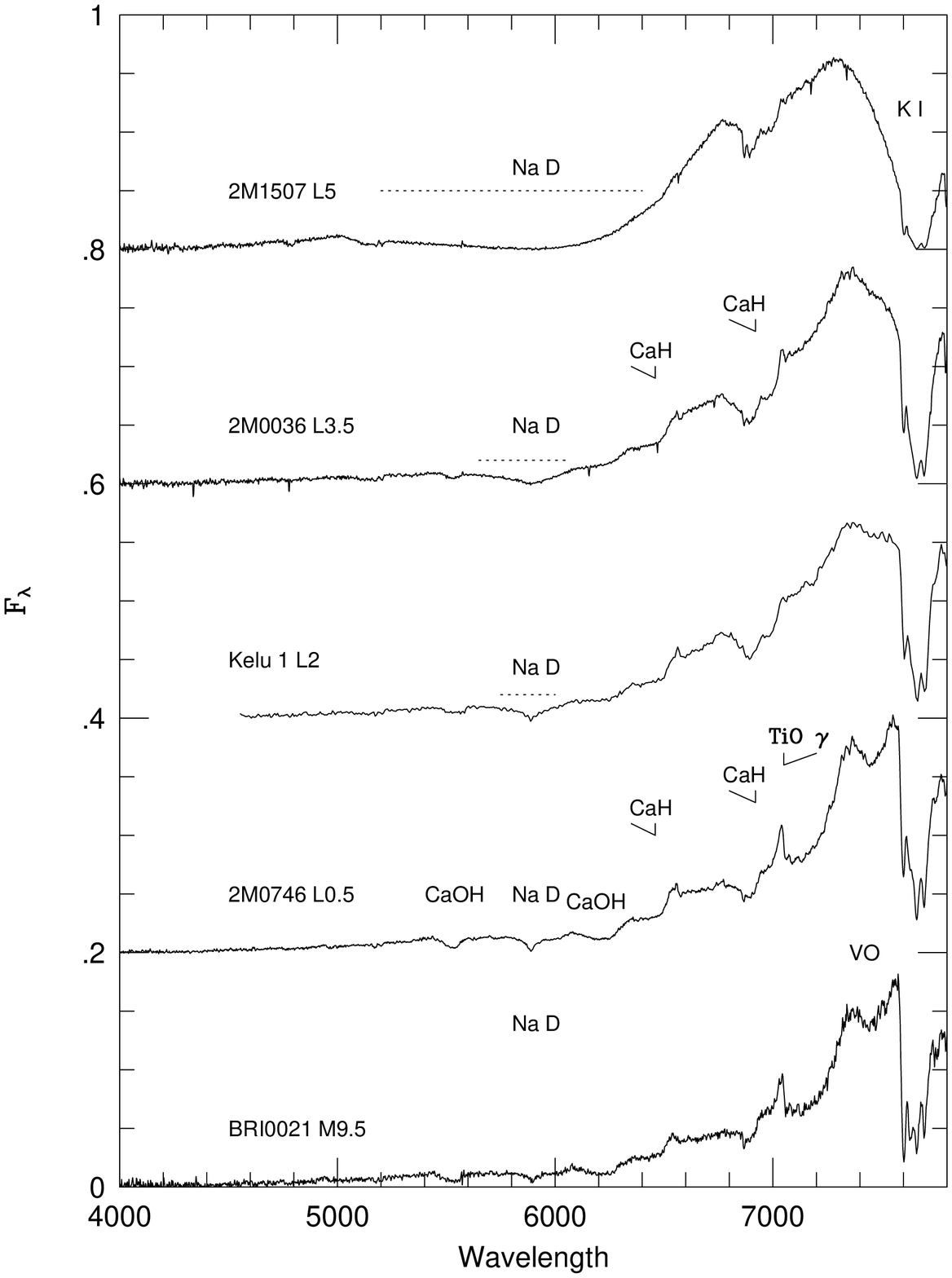}
\caption{}
\end{figure}

\begin{figure}
\plotone{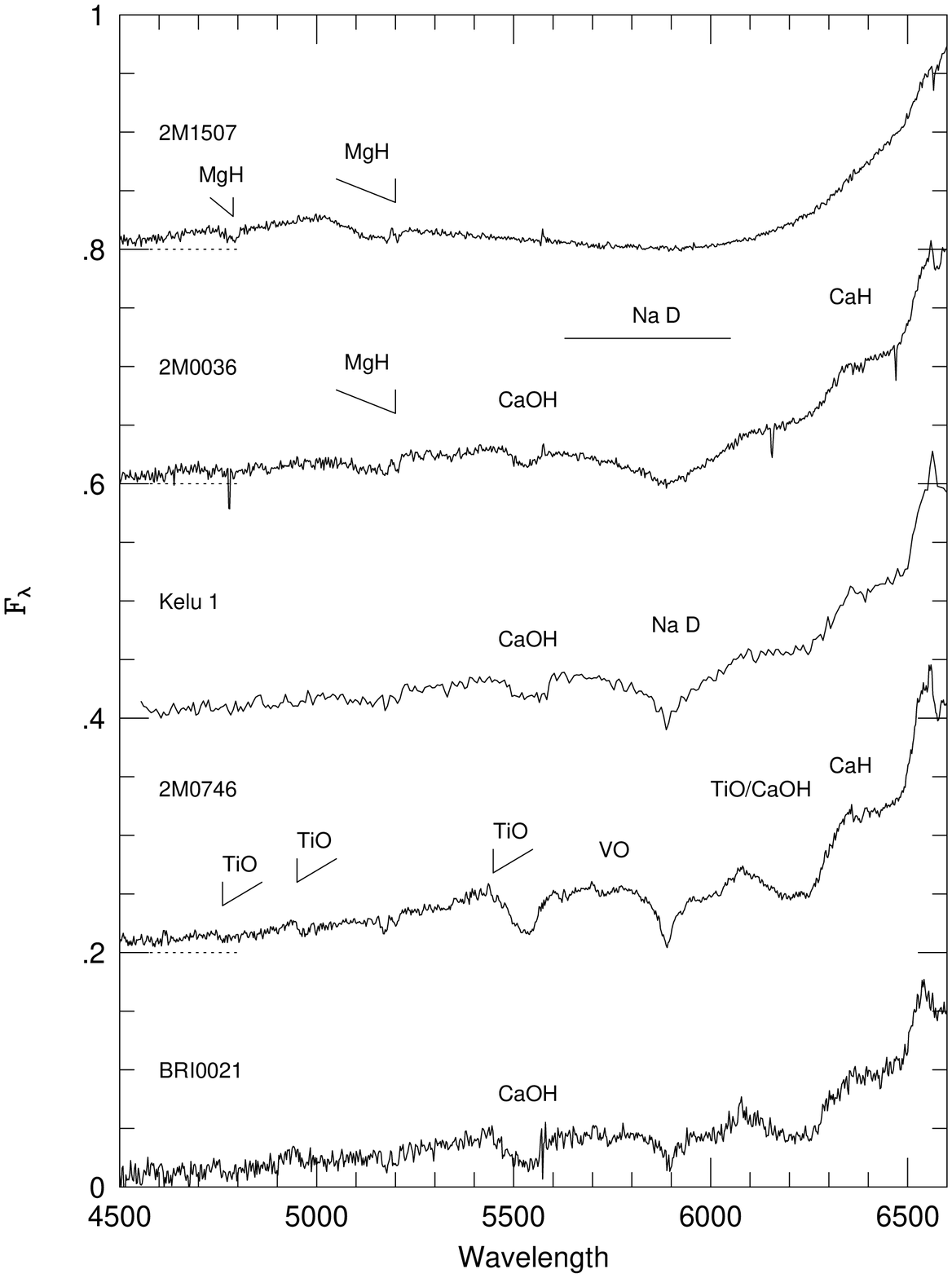}
\caption{}
\end{figure}

\begin{figure}
\plotone{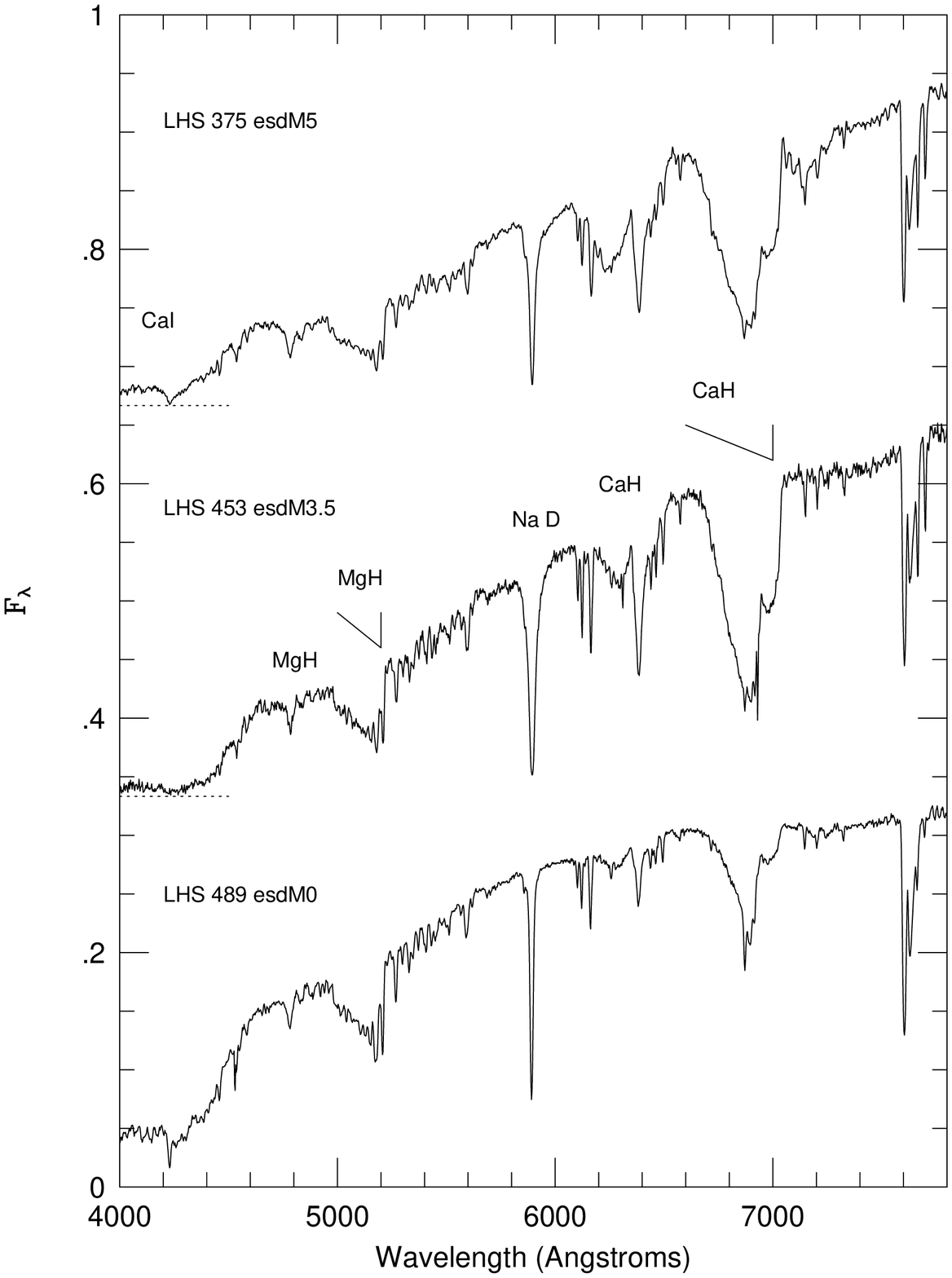}
\caption{}
\end{figure}

\begin{figure}
\plotone{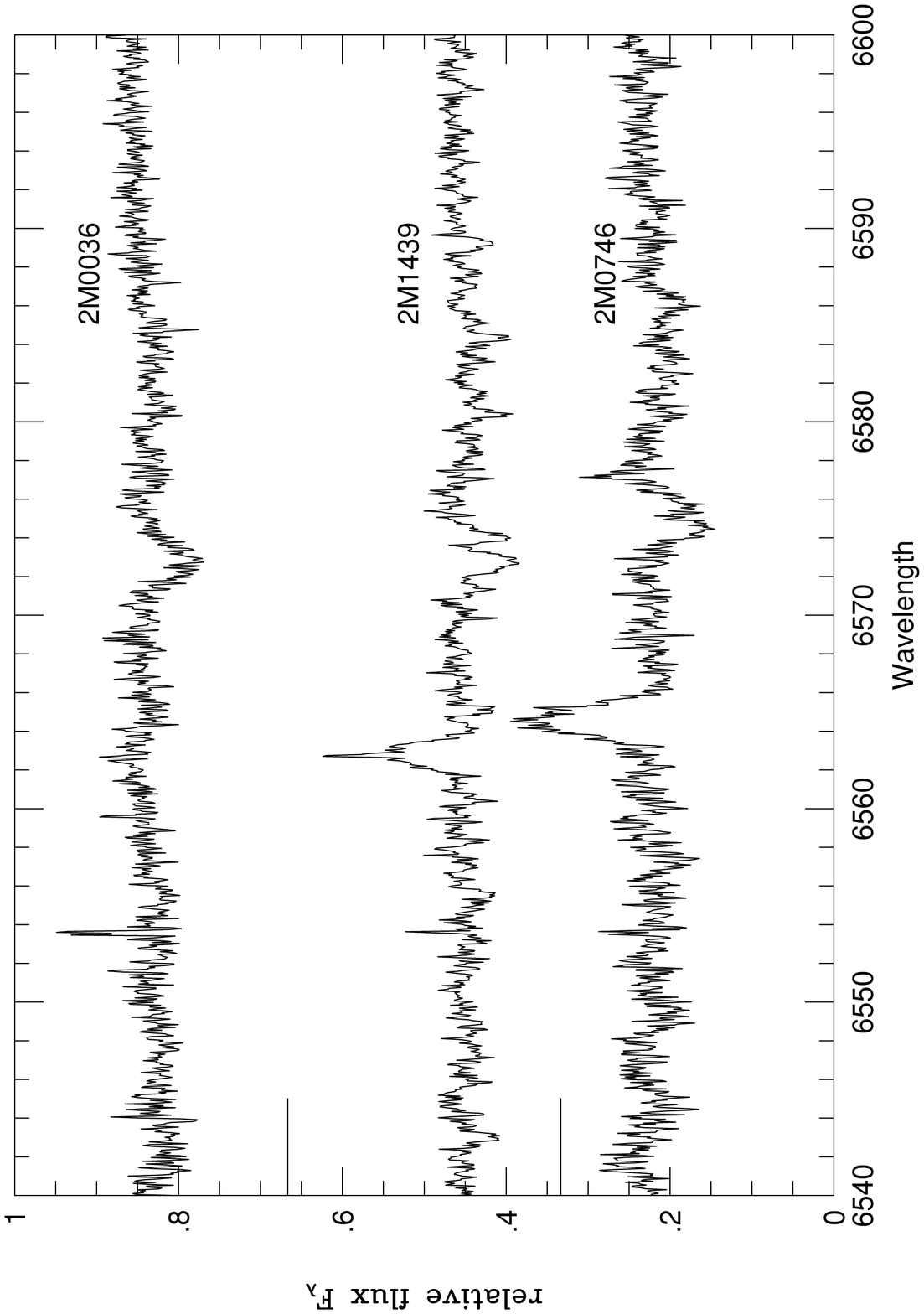}
\caption{}
\end{figure}

\begin{figure}
\plotone{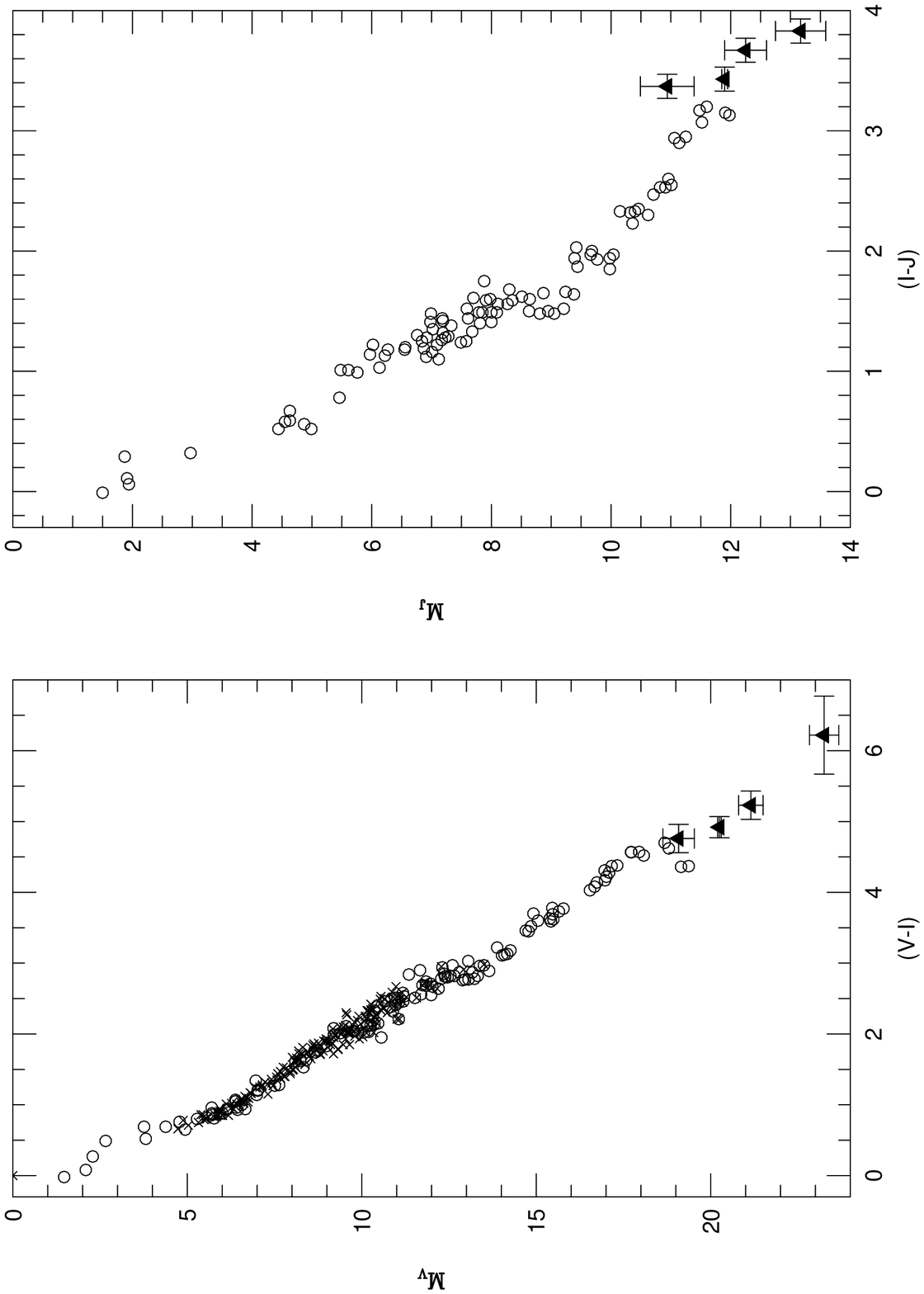}
\caption{}
\end{figure}

\begin{figure}
\plotone{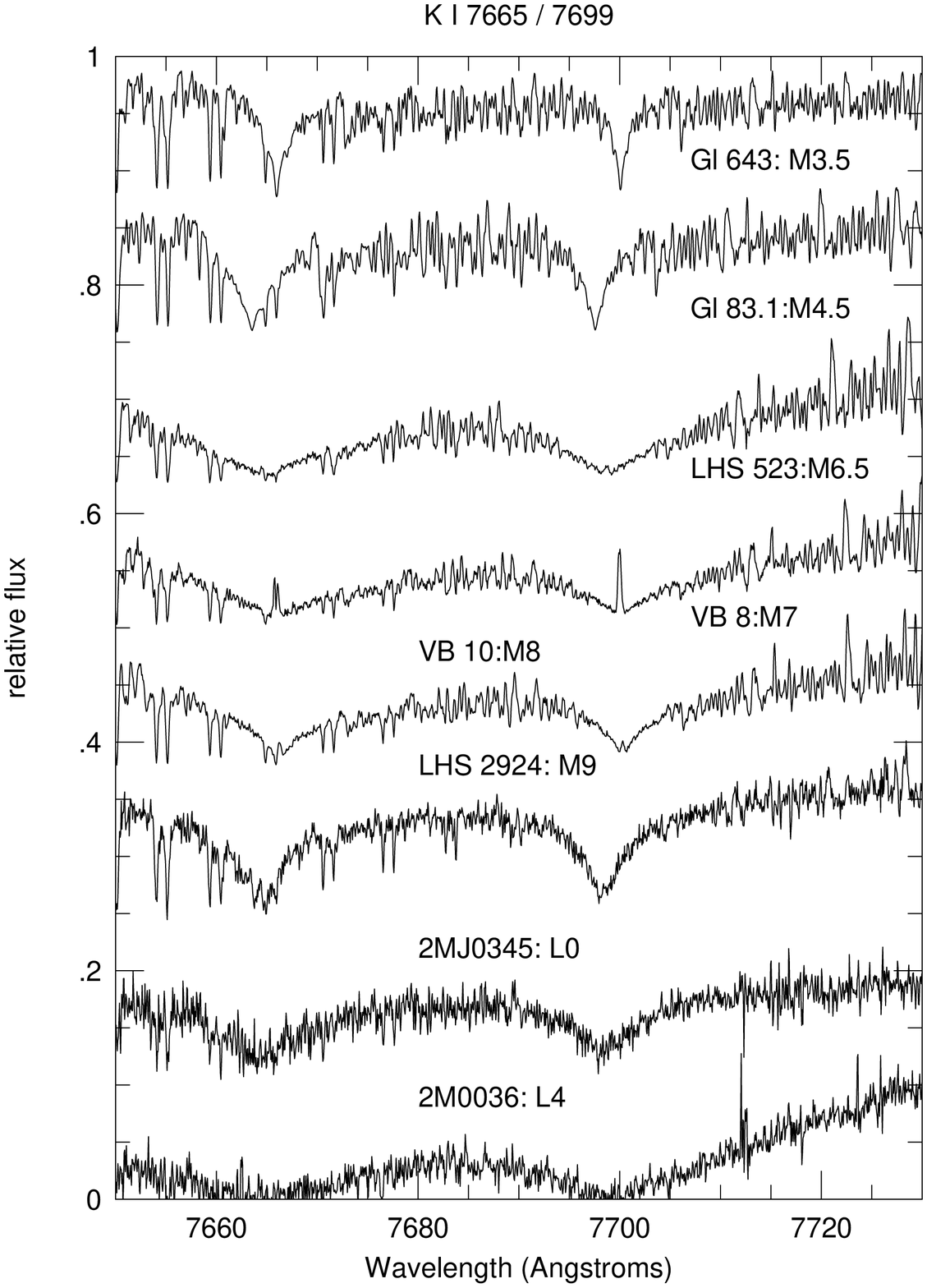}
\caption{}
\end{figure}


\begin{thebibliography}{}

\bibitem[Baraffe et al, 1998] {bcah98} Baraffe, I., Chabrier, G., Allard, F., Hauschildt, P.H. 1998, \aap, 337, 403
\bibitem[Bessell, M.S., 1979] {bes79} Bessell, M.S. 1979, \pasp, 91, 589
\bibitem[Bessell, M.S. 1989] {bes89} Bessell, M.S. 1990, \aaps, 83, 357
\bibitem[Boeshaar, 1976] {b76} Boeshaar, P. C. 1976, Ph. D. thesis, Ohio State Univ.
\bibitem[Burrows et al, 1997] {bur97} Burrows, A., Marley, M., Hubbard, W.B., Lunine, J.I., 
Guillot, T., Saumon, D., Freedman, R., Sudarsky, D., Sharp, C. 1997, \apj, 491, 856
\bibitem [Burrows \& Sharp, 1998] {bs99} Burrows, A., Sharp, C. 1999, \apj, 512, 843
\bibitem[Cottrell, 1978] {c78} Cottrell, P.L., 1978, \apj, 228, 544
\bibitem[Delfosse et al, 1997] {del97} Delfosse, X., Tinney, C.G., Forveille, T., Epchstein, N., Bertin, E.,
  Borsenberger, J., Copet, E., de Batz, B., Fouque, P., Kimeswenger, S., Le Bertre, T., 
Lacombe, F., Rouan, D., Tiphene, D. 1997, \aap, 327, L25
\bibitem[Epchtein et al, 199] {ep94} Epchtein, N., De Batz, B., Copet, E. et al 1994, in {\sl Science with
Astronomical Near-infrared Sky Surveys}, ed. N. Epchtein, A. Omont, B. Burton, P. Persei,
(Kluwer, Dordrecht), p. 3
\bibitem[Fegley \& Lodders]{fg96} Fegley, B., Lodders, K. 1996, \apjl, 472, L37 
\bibitem[Gizis, 1997]{g97} Gizis, J.E. 1997, \aj, 113, 806
\bibitem[Gizis et al, 1999]{g99}  Gizis, J.E., Monet, D.G., Reid, I.N., Williams, R. 1999, \aj, submitted
\bibitem [Hamuy et al, 1994]{ham94} Hamuy, M., Suntzeff, N.B., Heathcote, S.R., 
Walker, A.R., Gigoux, P., Phillips, M.M. 1994, \pasp, 106, 566
\bibitem [Hawley et al, 1996]{h96} Hawley, S.L., Gizis, J.E., Reid, I.N. 1996, \aj, 112, 2799
\bibitem[Kirkpatrick et al, 1999] {k99} Kirkpatrick, J.D., Reid, I.N., Liebert, J., Cutri, R.M.,
Nelson, B., Beichmann, C.A., Dahn, C.C., Monet, D.G., Gizis, J., Skrutskie, M.F. 1999a,
\apj, 519, 802 (paper I)
\bibitem[Kirkpatrick et al, 1999b]{k99b} Kirkpatrick, J.D., Reid, I.N., Gizis, J.E., 
Burgasser, A.J., Liebert, J., Monet, D.G., Dahn, C.C., Nelson, B. 1999, \apj, submitted
\bibitem[Leggett, 19992]{l92} Leggett, S.K. 1992, \apjs, 82, 351
\bibitem[Leggett et al] {l98} Leggett, S.K., Allard, F., Berriman, G., Dahn, C.C.,
Hauschildt, P. 1996, \apjs, 104, 117
\bibitem[Leggett et al, 1999] {l99} Leggett, S.K., Toomey, D.W., Geballe, T.R.,
Brown, R.H. 1999, \apj, 517, L139
\bibitem[Liebert et al, 1999] {lieb99} Liebert, J., Kirkpatrick, J.D., Reid, I.N., Fisher, M.D
1999, \apj, 519, 345
\bibitem[Lodders, 1999] {lod99} Lodders, K. 1999, \apj, 519, 793
\bibitem[]{a24} Magazzu, A., Martin, E.L., Rebolo, R. 1993, \apj, 404, L17
\bibitem[]{mb89} Marcy, G.W., Benitz, K.J. 1989, \apj, 344, 441
\bibitem[Monet et al, 1992] {mon92} Monet, D.G., Dahn, C.C., Vrba, F.J., Harris, H.C., 
Pier, J.R., Luginbuhl, C.B., Ables, H.D., 1992, \aj, 103, 638
\bibitem[Oke et al, 1995]{o95} Oke, J. B., Cohen, J. G., Carr, M., Cromer, J., Dingizian, A., 
Harris, F. H., Labreque, S., Lucinio, R., Schaal, W., Epps, H., Miller, J. 1995,
 PASP, 107, 375
\bibitem[Oppenheimer et al, 1998]{o98} Oppenheimer, B.R., Kulkarni, S.R., Matthews, K., 
van Kerkwijk, M.H. 1998, \apj, 502, 932
\bibitem[Pearse \& Gaydon, 1965] {pg65} Pearse, R.W.B., Gaydon, A.G. 1965, The Identification of Molecular Spectra,
(Chapman \& Hall, London)
\bibitem[Pesch, 1972] {p72} Pesch, P. 1972, \apj, 174, L155
\bibitem [Rebolo et al, 1992] {r92} Rebolo, R., Mart\'in, E.L., Magazzu, A.
1992, \apj, 389, L83
\bibitem[Reid \& Gilmore, 1984] {rg84} Reid, I.N., Gilmore G.F. 1984, \mnras, 206, 19
\bibitem[Reid \& Gizis, 1997] {rg97} Reid, I.N., Gizis, J.E. 1997, \aj, 113, 2246
\bibitem[Ruiz et al, 1997] {ru97} Ruiz, M.T., Leggett, S.K., Allard, F. 1997, \apj, 491, L107
\bibitem[Skrutskie et al, 1997] {sk97} Skrutskie, M.F. et al 1997, in {\sl The Impact of Large-Scale
Near-IR Sky Survey}, ed. F. Garzon et al (Kluwer:  Dordrecht), p. 187
\bibitem[Soderblom, 1990] {s90} Soderblom, D.R. 1990, \aj, 100, 204
\bibitem[Tinney et al, 1993] {ti93} Tinney, C.G., Mould, J.R., Reid, I.N. 
1993, \aj, 105, 1045
\bibitem[Tinney, 1996] {ti96} Tinney, C.G. 1996, \mnras, 281, 644
\bibitem[Tsuji et al, 1996a] {t96a} Tsuji, T., Ohnaka, K., Aoki, W. 1996, \aap, 305, L1
\bibitem[]{a45} Vogt et al, 1994, S.P.I.E., 2198, 362

\end{thebibliography}
\end{document}